\documentclass[sigconf]{acmart}

\AtBeginDocument{%
  \providecommand\BibTeX{{%
    \normalfont B\kern-0.5em{\scshape i\kern-0.25em b}\kern-0.8em\TeX}}}

\copyrightyear{2021}
\acmYear{2021}
\setcopyright{rightsretained}
\acmConference[UIST '21]{The 34th Annual ACM Symposium on User
Interface Software and Technology}{October 10--14, 2021}{Virtual Event,
USA}
\acmBooktitle{The 34th Annual ACM Symposium on User Interface
Software and Technology (UIST '21), October 10--14, 2021, Virtual Event,
USA}\acmDOI{10.1145/3472749.3474812}
\acmISBN{978-1-4503-8635-7/21/10}

\usepackage{subfig}
\usepackage{tabularx}
\newcolumntype{Y}{>{\centering\arraybackslash}X}

\newif\ifsubmit
\submitfalse
\ifsubmit
    \newcommand{\mingyuanz}[1]{}
    \newcommand{\leebird}[1]{}
    \newcommand{\liyang}[1]{}
    \newcommand{\peggy}[1]{}
\else
    \newcommand{\mingyuanz}[1]{{\textbf{\color{blue}{Mingyuan: {#1}}}}}
    \newcommand{\leebird}[1]{{\textbf{\color{green}Gang: {#1}}}}
    \newcommand{\liyang}[1]{{\textbf{\color{red}{Yang: {#1}}}}}
    \newcommand{\peggy}[1]{{\textbf{\color{orange}Peggy: {#1}}}}
\fi

\newcommand{\eg}{e.g.,\ }
\newcommand{\ie}{i.e.,\ }

\begin{document}

\title{HelpViz: Automatic Generation of Contextual Visual Mobile Tutorials from Text-Based Instructions}

\author{Mingyuan Zhong}
\authornote{This work was completed while the author was an intern at Google Research.}
\affiliation{%
  \institution{University of Washington}
  \city{Seattle}
  \state{WA}
}
\email{myzhong@cs.washington.edu}

\author{Gang Li}
\affiliation{%
  \institution{Google Research}
  \city{Mountain View}
  \state{CA}
}
\email{leebird@google.com}

\author{Peggy Chi}
\affiliation{
  \institution{Google Research}
  \city{Mountain View}
  \state{CA}
}
\email{peggychi@google.com}

\author{Yang Li}
\affiliation{%
  \institution{Google Research}
  \city{Mountain View}
  \state{CA}
}
\email{liyang@google.com}


\begin{abstract}
  We present HelpViz, a tool for generating contextual visual mobile tutorials from text-based instructions that are abundant on the web. HelpViz transforms text instructions to graphical tutorials in batch, by extracting a sequence of actions from each text instruction through an instruction parsing model, and executing the extracted actions on a simulation infrastructure that manages an array of Android emulators. The automatic execution of each instruction produces a set of graphical and structural assets, including images, videos, and metadata such as clicked elements for each step. HelpViz then synthesizes a tutorial by combining parsed text instructions with the generated assets, and contextualizes the tutorial to user interaction by tracking the user's progress and highlighting the next step. 
  Our experiments with HelpViz indicate that our pipeline improved tutorial execution robustness and that  participants preferred tutorials generated by HelpViz over text-based instructions. HelpViz promises a cost-effective approach for generating contextual visual tutorials for mobile interaction at scale.
\end{abstract}



\maketitle

\section{Introduction}
Mobile phones today provide a myriad of functionalities for a variety of everyday tasks. However, it is not always clear to a user how to access specific functionalities, such as dimming the screen or enabling tethering on the phone. It is thus common for mobile users to look for instructions on how to perform a specific task on the web.

Existing online tutorials are typically created by professionals or domain experts. One of the popular formats is a web document, which presents a sequence of steps where each step includes text instructions that are often accompanied by one or more images~\cite{chi2012mixt}.

To ease the generation of high-quality tutorials, previous work has proposed techniques to automate a creation process through demonstrations by expert users~\cite{grabler2009generating, chi2012mixt} or composing by a community~\cite{CommunityCommands_UIST09,CommunityVideos_CHI18}. However, the process of tutorial creation still needs a significant amount of human effort. As software is frequently updated, a tutorial needs to be constantly revised given that a task procedure might contain missing or mismatched details or its supporting screenshot images become outdated.

Thus, it is important to further reduce the cost of tutorial creation. In this work, we focus on utilizing the vast corpus of text instructions that are already available on the web. We tailor their content and presentation to the user's specific interaction context based on an understanding of the device infrastructure.
Prior work has suggested visualization techniques for tutorials that guide a mobile user through a task workflow by highlighting UI targets on a mobile device~\cite{wang2014evertutor}.
In addition, recent research demonstrated that by enhancing a text article with animated images, readers could better follow and comprehend the content~\cite{TextSlideshow_CHI20}.
Our goal is to convert text-based instructions to a visual tutorial that is seamlessly integrated into the user's interaction context on the mobile device.

\begin{figure*}
  \centering
  \includegraphics[width=0.9\linewidth]{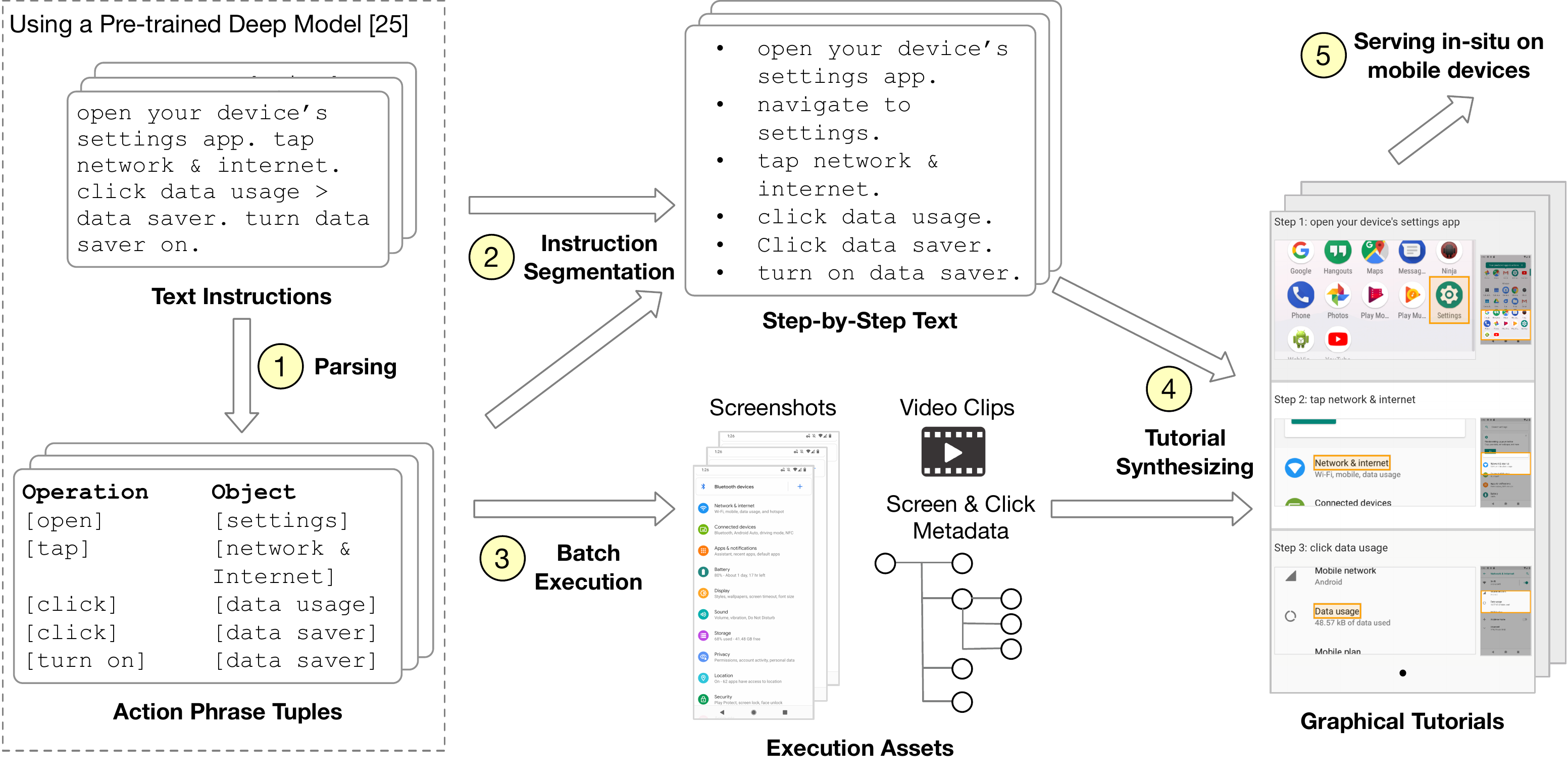}
  \caption{HelpViz converts text instructions that can be found abundantly on the web to visual tutorials and then serves these tutorials in the context of mobile interaction. (1)--(4): Our pipeline converts text instructions to visual tutorials. HelpViz generates the assets such as screenshots and interaction metadata of each instruction by actually executing it on a simulation infrastructure. It then synthesizes visual tutorials by combining these assets with segmented text instructions on a mobile device. (5): HelpViz Viewer serves visual tutorial in the interaction context of the user.}
  \Description{HelpViz overview. Step 1: text instructions are parsed into action phrase tuples. Step 2: text instructions are segmented into step-by-step text with the help of action tuples. Step 3: action tuples are used in batch exection to generate execution assets, including screenshots, video clips, screen and click metadata. Step 4: step-by-step text is combined with execution assets to synthesize graphical tutorials. Step 5: tutorials are served in-situ on mobile devices.}
  \label{fig:overview}
\end{figure*}

We present HelpViz, a tool that automatically transforms a batch of text instructions to contextual visual tutorials that can be interactively navigated on a mobile device (see Fig.~\ref{fig:overview}). Our pipeline (1) parses the instructional text to extract actions described in each step using a pre-trained deep model~\cite{li2020mapping}, (2) segments text instructions into steps based on the extracted actions, (3) executes the actions in a simulation infrastructure that manages an array of Android emulators, (4) synthesizes visual tutorials by combining segmented text and assets generated from execution, and finally (5) serves these tutorials in situ in the context of mobile interaction. Because such tutorials are equipped with screen assets and metadata captured from automatic execution of instructions, HelpViz is capable of tracking a user's progress and automatically anchoring the tutorial to the current step.

Our evaluation with 10 participants indicated that HelpViz received positive feedback from participants, for its rich graphical representation and progress tracking features, over text-based instructions. Our paper makes the following contributions:

\begin{itemize}
    \item The first pipeline that automatically generates visual tutorials for mobile tasks from raw text instructions, which can be those directly crawled on the web. Such a synthesized tutorial is served in the specific context of user interaction on the mobile device that automatically advances the focus of the tutorial by matching the user's progress.

     \item A set of algorithmic techniques and interaction design to address errors and incompatibility from automatic tutorial generation based on an AI model: \emph{beam-search} that addresses inaccuracy of the instruction parsing model and related interaction design to present alternatives in a tutorial, and \emph{look-ahead} that addresses the incompatibility between an instruction and an emulator during batch execution---these manifest how the imperfection of an automatic process can be addressed by closely coupling algorithmic and interaction design. These techniques were shown to substantially improve the robustness of tutorial synthesis.
     \item A set of experiments and findings that investigate the feasibility of visual tutorial generation and its impact on end-user experiences.
\end{itemize}

\section{Related Work}

Our work is related to three research areas in the literature: tutorial creation, contextual tutorials, and the general topic of incorporating an AI model in an interactive system.

\subsection{Structural Tutorial Creation}

The creation of tutorials for software has traditionally been completed by either user demonstration~\cite{grabler2009generating, chi2012mixt, eckhoff2018tutar, Torta_UIST17} or crowdsourcing~\cite{10.1145/2556288.2556986, GordonVLHCC2015codepourri}. Tutorials generated by \emph{demonstration} usually require tracing a user’s interaction flow while recording a screencast video of the user interface. Given a demonstration, MixT detects the most relevant areas in a step, and generates both screenshots and edited videos in a mixed-media document~\cite{chi2012mixt}. 
Torta allows creators to specify validation criteria for specific steps through file matching or explicit scripting~\cite{Torta_UIST17}. PPTutorial adds overlays on UI elements to facilitate locating them~\cite{li2015intelligently}.
On the other hand, \emph{crowdsourcing} can be an effective method for tutorial creation with the advantage of better scalability. ShowMeHow presents command strings in different applications and utilizes the crowd to match similar functionalities in order to translate commands between apps~\cite{showmehow}. Codepourri enables crowd learners to collaboratively annotate each step in a tutorial for a piece of code~\cite{10.1145/2556288.2556986}, while ToolScape allows learners to apply step-by-step annotations to instructional videos~\cite{GordonVLHCC2015codepourri}.

Previous approaches for tutorial creation commonly require human participation, in the form of user demonstration or crowdsourcing. In contrast, HelpViz automatically converts text instructions that exist on the web to visual tutorials by parsing and executing these instructions. Instead of logging the steps of a human user, or requiring manual collection and editing of assets, our system automatically gathers necessary assets generated from the execution of instructions, and synthesizes visual tutorials by combining the assets and text instructions.

\subsection{Context-Aware Tutorial Following}

The analysis of task executions allows users to discover relevant or more efficient workflows. Delta visualizes software workflows at multiple granularities and allows users to compare the trade-offs between them~\cite{kong2012delta}. Collective navigational traces within an instructional video have been used to facilitate the navigation within a tutorial~\cite{DataDriven_UIST14}. To recommend new workflows, Wang et al.~\cite{CommunityVideos_CHI18} extracted workflow topics and frequent command patterns. Khan et al. personalized the recommendations by utilizing web documentations and a user's command history~\cite{10.1145/2678025.2701387}.
Recent work has shown that by presenting its two-level hierarchy, a tutorial can be more easily consumed, especially for longer sequences~\cite{MakeupBreakdown_CHI21}.

In order to understand the current context, existing systems typically utilize accessibility APIs~\cite{RePlay_CHI19, rodrigues2018aidme} or image-based analysis~\cite{pongnumkul2011pause, wang2014evertutor} to determine the user action and match it against the tutorial. Interaction designs that previous work often use for desktop applications include direct overlays~\cite{kelleher2005stencils} and tooltips~\cite{grossman2010toolclips}, which can be linked with relevant images or videos. Previous work also focused on enhancing video-based tutorials, \eg pausing video when the user fell behind~\cite{pongnumkul2011pause} or presenting relevant videos to the user task~\cite{RePlay_CHI19}. Torii~\cite{head2020composing} enables flexible programming tutorials by linking code copies and edits with a reference implementation, preserving the context for code snippets. InterTwine~\cite{InterTwine_UIST14} bridges the web browser and desktop application through an explicit link, enabling shared context between the applications and easier help seeking.

On mobile systems, EverTutor~\cite{wang2014evertutor} detects user context through matching screenshots with pre-recorded templates. AidMe~\cite{rodrigues2018aidme} allows voice-based tutorial creation and playback tailored to visually impaired users through an accessibility service. Similar to previous work, a tutorial generated by HelpViz is presented to a user contextually. Specifically, HelpViz utilizes the accessibility APIs to match the user's current screen with a step in the tutorial. Importantly, the assets and metadata that enable such context awareness are automatically captured by the HelpViz pipeline, without any human effort.

\subsection{Incorporating AI Models into Interactive Systems}
    
Our work is broadly related to the topic of incorporating the output of an AI model into an interaction task. Because the imperfection of the instruction parsing model or the inherent mismatch between an instruction and an emulator, there can be errors for generating tutorial assets during the automatic execution. Several strategies have emerged in the literature for integrating an AI model. Previous work has proposed interactive machine learning that iteratively refines an AI model using user input (e.g.,~\cite{fails2003interactive, fogarty2008cueflik}). Previous work has also developed interaction techniques to allow users to easily navigate alternative model results (e.g.,~\cite{textentry, 10.1145/1449715.1449724}). 

In our task, we aim to generate visual tutorials from a batch of text instructions in parallel, without human intervention. Our intended users are help seekers who are likely unfamiliar with the issues themselves. As a result,
interactively refining tutorial generation is not feasible in HelpViz. Instead, we choose to reduce tutorial generation failures by capturing model and emulator execution uncertainty, via beam search~\cite{meister2021beam} and look-ahead, and then presenting the captured uncertainty and alternatives to the user via appropriate interaction designs.

\section{HelpViz}
HelpViz consists of a pipeline that converts text instructions to visual tutorials (see Figure~\ref{fig:overview}.1--\ref{fig:overview}.4), and a mobile tool, HelpViz \textit{Viewer}, that serves a visual tutorial in the interaction context of the user (Figure~\ref{fig:overview}.5).
The web offers a rich corpus of text instructions, \eg online help documents for how to use Android\footnote{Android Help, \url{https://support.google.com/android}} or iOS\footnote{iPhone - Official Apple Support, \url{https://support.apple.com/iphone}}. For example, to use less mobile data, users can find instructions that guide them through a workflow: \emph{``Open your device’s settings app. Tap network \& internet. Click data usage > data saver. Turn data saver on.''}
It requires a user matching the information with their device screen, one action at a time.
To generate visual tutorials for such instructions with useful graphical assets, we propose an automatic approach to parse the text instructions, execute the actions on a device emulator, and interactively present on user's device. Below we describe the components in our pipeline.

\subsection{Instruction Parsing \& Segmentation}
We first use an instruction parsing model to extract the action phrase tuples that describe each \textit{step} of the text \textit{instruction}\footnote{In this paper, an \textit{instruction} refers to a text description of a sequence of steps that complete a task, while a \textit{step} refers to a single executable action from the instruction.}, using a pre-trained deep model~\cite{li2020mapping}. The model takes the raw text of an instruction as input. It outputs a sequence of action tuples, which consists of the operation (\eg ``click'', ``turn on'') and target object (the element on which to apply the operation) at each step (see Fig.~\ref{fig:overview}.1). 

The extracted action tuples from instruction parsing are used in two separate processes, instructional text segmentation and automatic execution. We split the raw text of an instruction into multiple steps by matching each phrase tuple to its respective sentence(s) in the instruction (see Fig.~\ref{fig:overview}.2). Our pipeline generates visual assets for each step of the task by executing the extracted actions on an Android emulator, similar to prior work that mines UI layouts~\cite{ERICA_UIST16,Rico_UIST17}.

\subsection{Batch Execution on Emulators \& Tutorial Synthesizing}\label{sec:execution}

We developed an infrastructure that can automatically execute a batch of parsed instructions in parallel (see Fig.~\ref{fig:overview}.3), by running each instruction on a separate emulator, which is a standard Android Emulator spawned on a distributed computing platform controlled through ADB commands (e.g., sending a tap command). Given an action tuple extracted from an instruction, our instruction \textit{executor} identifies and acquires a target element on the current UI screen of the emulator, which triggers a transition to the next screen. However, it is likely that the target element described in the tuple cannot be immediately located, \eg when the screen is still loading or the element is outside of the screen. The executor will either wait for the screen to be loaded or attempt to scroll the screen to find the element. Note that these dynamic aspects are not addressed in previous work~\cite{li2020mapping}, which assumes a static setup for supervised sequence modeling.

While executing an instruction, the executor collects a rich set of assets for each action taken, including UI screenshots, screencast video clips that capture transitional UI animations, and metadata such as UI view hierarchies and elements being clicked. These assets are necessary for synthesizing visual tutorials and enabling step matching to serve a tutorial contextually that we will describe in the following sections. Our pipeline then combines the assets captured by the executor and the segmented instruction steps from the previous stage to compose visual tutorials (see Fig.~\ref{fig:overview}.4). HelpViz structures a tutorial as a sequence of steps where each step consists of text instructions and its corresponding screenshot overview and close-up view as well as a video segment. 

There are two challenges in automatically executing instructions crawled from the web on an emulator. One is due to the inaccurate inference results of the parsing model. The other issue is rooted in the mismatch between the instruction and the actual emulator interfaces. Both issues can cause failures for automatic execution. We address these issues as following.

\subsubsection{Addressing Model Parsing Errors} 

By default, the instruction parsing model~\cite{li2020mapping} extracts a best sequence of operation-object tuples for actions from the text instructions by greedy decoding. However, with the presence of multiple steps in an instruction, such greedy decoding may generate a sub-optimal sequence if there are errors in the decoded actions, because for each step it only selects the best action depending on previous actions, which can be a local optimum considering the full sequence. To address this issue, we use \textit{beam search} to track the top $k$ decoded sequences at each step to improve the overall parsing result. 

Specifically, the instruction parsing model parses an instruction and generates multiple possible action sequences (beams). In generating the action sequences, the model aims to maximize the probability of the overall action sequence, rather than the probability of individual actions. In our experiments, we chose to execute the top 3 sequences (\ie $k = 3$) of an instruction. Thus, our emulator infrastructure launches three emulator instances for each instruction, and each of the decoded sequences is executed independently on a separate emulator instance. To merge the execution results into a consistent form that can be served to the user, HelpViz first discards the results from the executions that cannot reach the final step. Then, for executions that reach the final step, HelpViz merges the steps shared by executions and preserves steps that differ as alternatives. Finally, HelpViz sorts these alternatives based on the decoding score of the sequence they belong to.
This approach allows the model to output several slightly different parsing results and reduces the likelihood that the greedy-decoded sequence contains errors and leads an dead end during execution.

Although beam search is a common technique in decoding models, it has special implications in our case for both system and interaction design. For the system design, our executor needs to handle multiple alternatives of executing an instruction by spawning multiple threads and then merging the results of the execution of these threads, as discussed above. For the interaction design, we need to design appropriate interfaces for the user to examine these alternatives, which we will discuss later.

\subsubsection{Addressing Instruction-Emulator Mismatch}

It is possible that an instruction does not exactly match the app UIs on the emulator. For example, the instruction can be outdated as the UI of an app running on the emulator has been recently updated. In order to mitigate the issue, we implemented \textit{look-ahead} in the executor to maximally execute an instruction by skipping steps that cannot be realized. Specifically, the executor allows a maximum of five unsuccessful actions (tapping or scrolling) for each task. If the executor still could not perform the action described in the current step after five attempts (\ie the target element in the current step does not appear on the screen), it then skips the current step, looks for the target element described in the next step, and attempts to execute the next step's action if found. For example, an instruction may assume that an expandable view is folded and include a step of ``tap \textit{show all items}''. During execution, the view might have already been expanded on the emulator (resulting from a prior action or a difference in app versions) and skipping the corresponding step in the tutorial would be warranted. In this case, after failing to find the button ``\textit{show all items}'' after five scrolls, the executor would look for an item in the list that corresponds to the next action in the instruction.

\subsection{Serving a HelpViz Tutorial in Context}

\begin{figure*}[h]
  \centering
  \subfloat[A generated HelpViz tutorial served on an Android phone, with the first three steps visible.]{\frame{\includegraphics[width=0.28\linewidth]{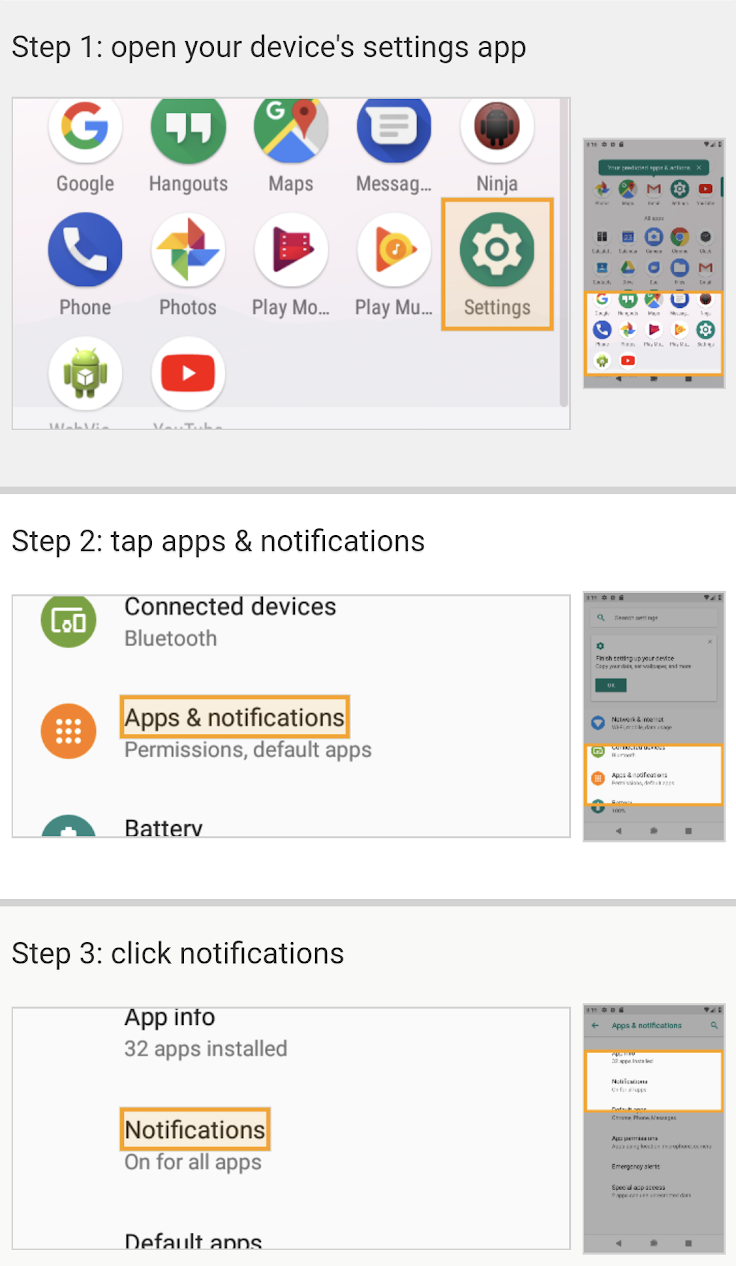} }}
  \qquad
  \subfloat[A user navigates to the \textit{Apps \& notifications} screen in \textit{Settings} on their device as described in tutorial Step~2.]{\frame{\includegraphics[width=0.28\linewidth]{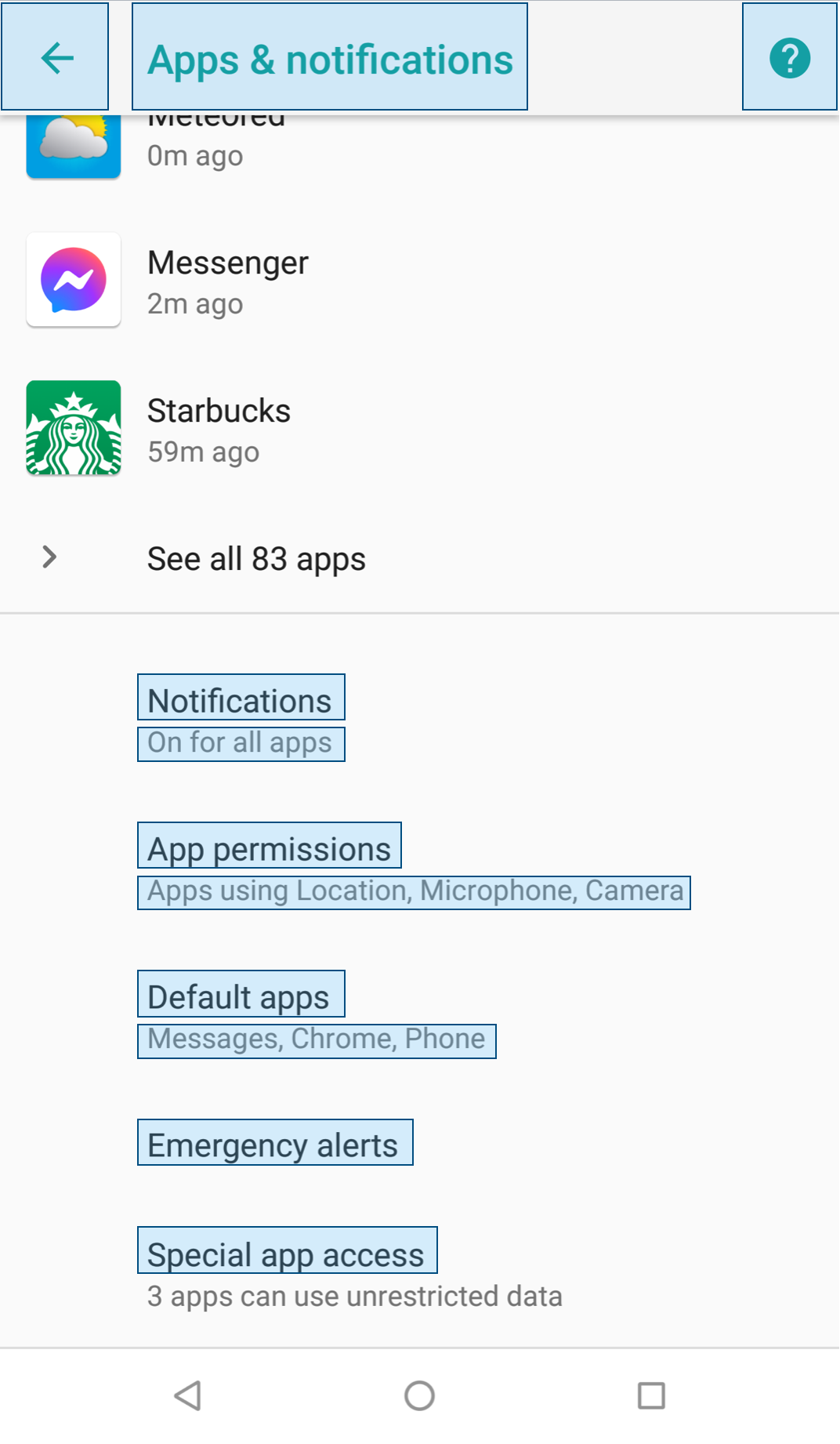} }}
  \qquad
  \subfloat[HelpViz highlights the next step after the user has executed the first two steps.]{\frame{\includegraphics[width=0.28\linewidth]{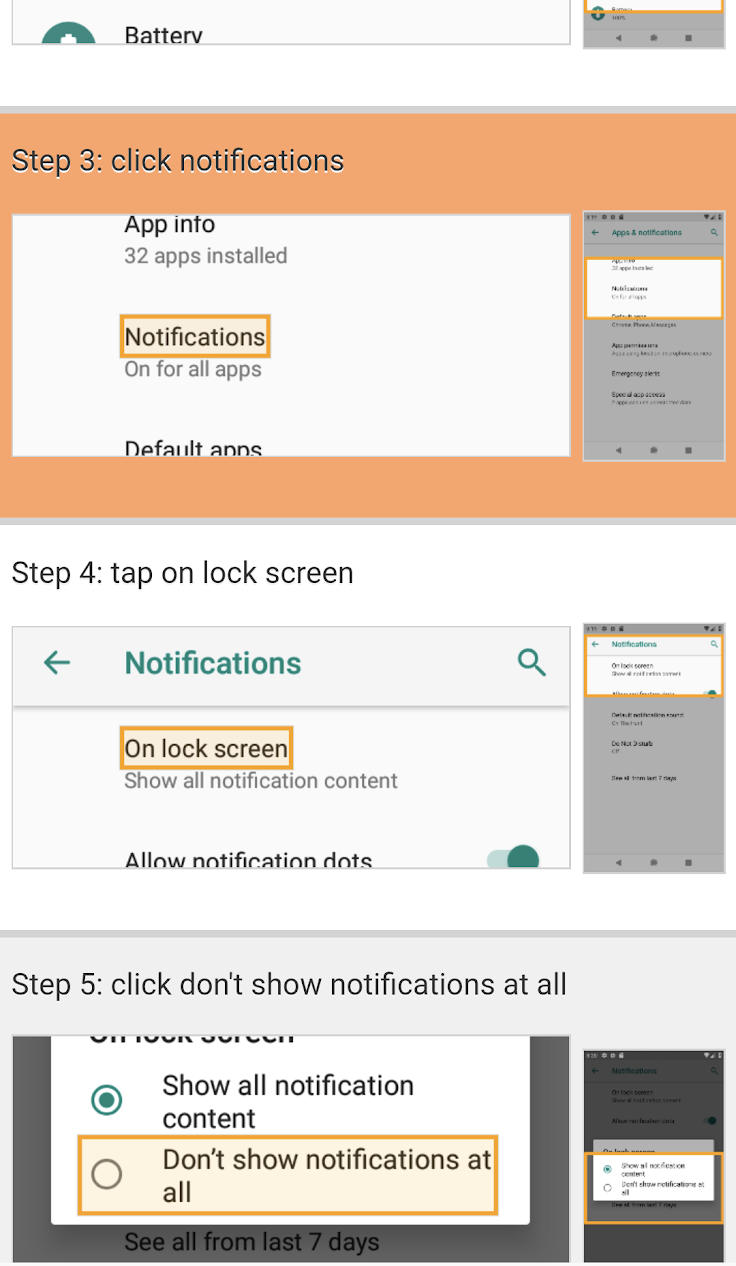} }}
  \caption{HelpViz Viewer serves an automatically-generated visual tutorial in the context of user interaction. In this example, the tutorial illustrates the task of ``turning off all notifications.'' The original text instruction is: ``Open your device's \textit{Settings} app. Tap \textit{Apps \& notifications}. Click \textit{Notifications}. Tap \textit{On lock screen}. Click \textit{Don't show notifications at all}.'' (Formatting added by the authors for clarity.) Note that in (b), elements with text matched to Step 3 are highlighted to illustrate contextual matching. The matching allows HelpViz to scroll to and highlight that step, illustrated in (c). The blue boxes in (b) are \textit{not} present during user interaction.}
  \Description{Screenshots of HelpViz tutorial. (a) HelpViz Viewer screenshot with the first three steps shown. The screen consists of a list showing the first three steps, each with a text description, a corresponding close-up view around the target object, and an overall screenshot. (b) Screenshot of Settings page "Apps & Notifications." Some elements are highlighted to show matching with a tutorial step. Some of the highlighted elements include the graphical back button, the page title "Apps & Notifications", and many list items, such as "Notifications", "App permissions", etc. (c) Screenshot of HelpViz Viewer. The next step (Step 3) is at the top of the screen with its background highlighted in orange. Also visible below are steps 4 and 5, with the same layout as in Figure 2.a.}
  \label{fig:demo}
\end{figure*}

To support users following instructions, we develop HelpViz \textit{Viewer} that serves a HelpViz visual tutorial directly on the mobile device where the user seeks help. In Fig.~\ref{fig:demo}a, the Viewer loads a generated HelpViz tutorial for turning off all notifications. Each step has two screenshot thumbnails: the close-up view is a cropped image around the target element, and the overview of the UI screen highlights the cropped area, similar to prior work~\cite{Overview_detail, chi2012mixt}. If any step requires scrolling to reach the target element, the thumbnail will be substituted with a short scrolling animation on that screen when the step is either scrolled to the top (including auto-scroll by HelpViz) or being tapped on. When multiple alternatives are available for a step, a user can swipe horizontally to switch between the alternatives.

To follow the instructions, a user switches to the target app and performs the actions described in the tutorial (Fig.~\ref{fig:demo}b). Once the user switches back to HelpViz for further help, HelpViz Viewer automatically scrolls to the step that the user is working on and highlights that step (Fig.~\ref{fig:demo}c). The highlight color automatically fades out within one second to avoid distracting the user.

To synchronize the current step in the tutorial with the actual progress of the user, HelpViz registers a background accessibility service to compare the UI elements on the app screen with the elements of each step in the tutorial. The accessibility service captures the current view hierarchy of the target app every time the screen updates by listening to system \textit{AccessibilityEvent}s and records text-related properties (\ie \textit{Text}, \textit{ContentDescription}, and \textit{HintText}) of each UI element. These texts are matched against the texts stored for each step in tutorial metadata using Jaccard index~\cite{jaccard1912distribution} (also known as Intersection over Union) as a similarity metric. 
Specifically, it is computed as the number of common tokens in both sets of text (\ie their \textit{intersection}) divided by the number of all unique tokens in the two sets (\ie their \textit{union}). To illustrate the matching result, Fig.~\ref{fig:demo}b highlights the elements with text matched to Step 3 of the tutorial.

In rare cases, when multiple dynamically generated elements are present on a screen, the accessibility service may fail to match the current screen with an extracted screen from the pipeline if text in most of these elements are different from what we captured during emulation. As a fail-safe, HelpViz remains at the last viewed step in this case.

\section{Pipeline Performance Experiments} \label{sec:generating_tutorials}

\begin{table*}
  \caption{Summary of Actions Executed and Assets Collected by our pipeline on processing the PixelHelp Dataset.}
  \renewcommand{\arraystretch}{1.5}
  \label{tab:stats}
  \begin{tabular}{cccccc}
    \toprule
    & \textbf{\# Actions Executed} & \textbf{\# Words} & \textbf{\# Images} & \textbf{Video (seconds)} & \textbf{\# UI Elements} \\
    \midrule
    Minimum & 4    & 11   & 4    & 70.1   & 64 \\
    Maximum & 28   & 48   & 21   & 573.8   & 471 \\
    Mean    & 10.7 & 18.0 & 8.2  & 252.9 & 173.8 \\
    Total   & 1997 & 3368 & 1528 & 47,296 & 32,494 \\
  \bottomrule
\end{tabular}
\end{table*}

The HelpViz pipeline robustly handles instructions crawled from the web. We converted the PixelHelp dataset\footnote{PixelHelp dataset~\cite{li2020mapping}: \url{https://github.com/google-research-datasets/seq2act}}, which contained 187 multi-step text instructions for using Android Pixel phones, to visual tutorials using the pipeline. The execution of each instruction completed in less than 15 minutes and all the executions could be performed in parallel. In our tests, we ran 50 emulators simultaneously to speed up tutorial generation. Including attempts from beam search and look-ahead, our pipeline executed 1997 actions in total and produced and gathered necessary assets (see Table~\ref{tab:stats} for a summary).

\begin{table}
  \caption{Effect of Beam Search (BS) and Look-Ahead (LH)}
  \renewcommand{\arraystretch}{1.5}
  \label{tab:bs_lh}
  \begin{tabular}{ccccc}
    \toprule
    & \textbf{Baseline} & \textbf{BS} & \textbf{LH} & \textbf{BS + LH} \\
    \midrule
    Steps Executed  & 2.82    & 3.17   & 2.94    & 3.35 \\
    Completion Rate & 67.1\%   & 76.6\%   & 69.1\%   & 80.9\% \\
    \bottomrule
\end{tabular}
\end{table}

In this experiment, we investigate whether beam search and look-ahead, the two techniques we employed in our pipeline for robust batch execution (Section~\ref{sec:execution}), improves the performance of our automatic instruction execution (see Table~\ref{tab:bs_lh}). To do so, we mainly look at two metrics: (1) \textit{steps executed}, defined as the average number of steps successfully executed of each instruction, and (2) \textit{completion rate}, defined as the proportion of successfully executed steps to the total number of steps in each instruction. For example, an instruction with 5 total steps has a completion rate of 60\% if 3 of its steps are successfully executed. We compare four configurations of the pipeline: (1) beam search only (\textit{BS}), (2) look-ahead only (\textit{LH}), (3) beam search with look-ahead (\textit{BS+LH}), and finally (4) the baseline condition that uses neither beam search nor look-ahead (\textit{Baseline}). 

We used 50 emulators in parallel to execute the 187 instructions for each configuration during the experiments. Thus, there were 4 batches of execution to process all the instructions for the Baseline or LH conditions, and 12 batches for the BS or BS+LH conditions because they both used beam search with a beam size of 3, resulting in $187 \times 3$ sequences to execute. This study evidences why it was necessary to support parallel execution. On average, an execution of one instruction took 8 minutes. Even for the Baseline condition, sequential execution would result in an execution time of over 24 hours. The additional execution required for implementing beam search and the potential need to support different platform or app version further amplifies the need for parallel processing.

\textbf{Beam Search Only.}
Of the 187 generated tutorials, 32 tutorials (17.1\%) benefited from the extra beams, where the second or third beams executed more correct steps than the first beam. The first beams in these cases (\ie the baseline condition) averaged only 1.22 steps executed, while the beam selected by beam search average 3.25 steps executed, successfully executing two more steps on average. With beam search, the average completion rate for all instructions improved from 67.1\% of the baseline to 76.6\%. 

\textbf{Look-Ahead Only.}
We found that 10 tutorials (5.3\%) were improved by look-ahead (\ie containing at least one look-ahead step in any of its beams), increasing from an average of 2.4 steps executed of the baseline to 4.6 steps executed. Overall, there was also an increase in completion rate versus baseline (69.1\% vs. 67.1\%).

\textbf{Beam Search with Look-Ahead.}
We found that enabling beam search along with look-ahead improved the effectiveness of look-ahead, with look-ahead being applied to 41 tutorials (21.9\%), compared to only 10 with look-ahead alone. The average number of executed steps in these tutorials increase from 1.59 of the baseline to 3.46. Meanwhile, the addition of look-ahead also made beam search more effective, with the number of tutorials benefiting from beam search increasing from 32 for beam search only to 39 (20.9\%). 
Overall, the completion rate rose from 67.1\% of the baseline to 80.9\% of beam search with look-ahead. The number of successful steps executed increased from 2.82 to 3.35, an 18.8\% increase. In total, 63 tutorials (33.7\%) benefited from beam search, look-ahead, or both. The existing failure cases mostly occurred due to a substantial mismatch between the instruction text and the app interface caused by app updates. In addition, because the emulator was automatically log into some apps, some actions that require authentication failed.

\textbf{Summary.}
Our results demonstrate the synergy of adopting both beam search and look-ahead, showing substantial improvement over the baseline technique. While we leverage and adapt general techniques, our pipeline is novel for fully-automatic tutorial creation from text instructions. With this infrastructure, we can transform more text instructions to visual tutorials at scale.

\section{User Evaluation}

To investigate how mobile users receive visual tutorials generated by HelpViz in comparison with the original text instructions, we conducted a user study with 10 participants. We used the HelpViz tutorials generated from the PixelHelp dataset (see Section~\ref{sec:generating_tutorials}) for the experiments.

\subsection{Participants}

We recruited 10 participants from an IT company (2 female, 8 male). Each participant received a gift code worth \$15 US dollars for their completion of the study. 
All the participants had 5-year experience using a smartphone at minimum. Four participants rated their knowledge of using an Android phone as moderate and that they occasionally seek help; the remaining six participants rated themselves as experts and seldom seek help. While it would be preferable if the participants had little or no prior experience with Android, under the challenges of the COVID-19 pandemic when the research was conducted, each participant needed to use their own device in a remote setting. This limited us to recruiting participants who owned an Android phone, which meant that they were likely to have some experience with Android. Although these users may not rely on tutorials for day to day tasks, they still use tutorials when encountering unfamiliar or complex tasks.

\subsection{Apparatus}

For this remote study, we asked the participants to run the HelpViz study app on their personal Android phones. Participants used Android phones with a variety of versions, which six of the phones were running Android v10, two on Android v9, one on Android v8, and one on Android v6. Three participants used Google Pixel phones, two used Samsung Galaxy phones, one used a Motorola Moto phone, one used a OnePlus phone, while the others did not disclose their phone models. Of the seven phones that we know the model name, four had a screen size between 5.5 and 5.9 inches, and three were between 6.0 and 6.4 inches. We do not expect the variations in phone models and screen sizes to affect our study, and none of the participants reported issues regarding device compatibility.

We randomly selected 22 tutorials generated from the PixelHelp dataset, generated with beam search and look-ahead in Section~\ref{sec:generating_tutorials}.
Each task involved navigating to a certain page in an Android app. The apps used in the tasks consisted of four apps, Settings (8 tasks), Clock (2 tasks), Google Photos (4 tasks), and Google Chrome (6 tasks). The tasks required between three and seven steps to complete. The study app logged all user interaction with time stamps during the study. When users are seeking help from a tutorial, they usually do not have a good idea of the steps to take. Therefore, we did not include titles for the tutorials to reduce the likelihood of a participant familiar with the system guessing the steps without looking at the tutorials. 

\subsection{Experimental Design and Procedure} \label{sec:exp_procedure}

We employed a within-subjects design for this study, where the only factor was Technique (HelpViz Tutorial or Text-Only). We presented 11 tasks for each technique, with the first one used for warm-up. The presentation order of the two techniques was counterbalanced between participants. The assignment of tasks to each technique was also randomized across participants. We counterbalanced the tasks across techniques such that a task was presented the same number of times in each technique.

Our experiment consisted of three parts. First, we introduced the tasks to the participants and asked them to install the study app on their phones. Then, we asked the participants to complete a warm-up task and 10 tasks for each technique tested. We informed the participants to switch between the tutorial and the target app whenever necessary for completing a task. The participants were allowed to take a break after each task. Finally, we asked them to fill in a post-study survey about their experiences during the study and collected demographic information. Specifically, we asked them to what extent they agreed or disagreed with the following statements for each technique on the five-point Likert items below (\ie each on a scale of 1 to 5). We referred to their short names (in parentheses) in the Results section. Each participant's subjective ratings for the six items were summed to create an \textit{overall scale} on a range between 6 and 30 for each technique~\cite{boone2012analyzing}.

\begin{enumerate}
    \item ``It was easy to understand the instructions.'' (\textit{Understand})
    \item ``I could easily find the corresponding action.'' (\textit{Find action})
    \item ``I was able to complete the tasks without problems.'' (\textit{Complete task})
    \item ``It was faster to use tutorials like this to complete a task.'' (\textit{Speed})
    \item ``I was satisfied with tutorials like this.'' (\textit{Satisfaction})
    \item ``When finding help online, I would prefer to use tutorials like this.'' (\textit{Preference})
\end{enumerate}

We also collected background information, including the number of years participants had used a smartphone, daily smartphone usage in hours, and their knowledge of using an Android phone. Finally, we discussed with the participants on their impressions and experience while using both types of tutorials.

\subsubsection{Analysis procedure}
In order to test for potential differences in \emph{Completion Time} (\ie the time elapsed between trial start to reaching the target screen) and the \emph{Number of App Switches} (\ie the number of times a participant switched between the tutorial and the target app during a trial) for each technique, we performed a Mixed Model analysis of variance on each of the dependent variables, modeling \textit{Technique} as an independent factor and \textit{ParticipantID} as a random effect. 

In order to test for potential difference in participants' subjective ratings, we conducted a paired-samples \textit{t}-test on the overall Likert scale to compare the techniques. We also compared the ratings for each technique using the Wilcoxin signed-rank test. Finally, we summarized the participants' subjective feedback.

\subsection{Experimental Results}

The 10 participants completed a total of 200 trials. Due to the nature of remote studies where participants performed the tasks on their own devices in their own environment, we needed to remove 63 trials. Of these, 49 trials were excluded due to the mismatch of user devices or apps ("android fragmentation"). For example, participants' devices had photos apps from a different vendors, or Samsung's Settings app instead of Pixel's. Inherently, these were different apps instead of different versions of an app. These differences required vastly different tutorials, which could be easily addressed by letting HelpViz generate tutorials corresponding to these devices from instructions specific to these apps, or better screening participants for the remote study. The rest of the cases were due to exceptions in the study that are not related to HelpViz, \eg apps on participants’ devices restarting during the task. All these cases could have been easily avoided in a lab study where a designated device is used, which was not possible during the pandemic.
This resulted in a total of 137 trials used for analysis. 

\subsubsection{Performance}
We did not find statistically significant difference in the \textit{Completion Time} ($F_{1, 129} = 2.42, p \approx .12$) and the \textit{Number of App Switches} ($F_{1, 129} = 1.22, p \approx .27$) for \textit{Technique}, \ie HelpViz versus Text-Only. The mean \textit{Completion Time} was 33 seconds for HelpViz, and 30 seconds for the Text-Only instructions. The mean \textit{Number of App Switches} was 2.62 times for HelpViz, and 2.34 times for text instructions. To further test whether there was a difference between the two techniques, we also included the \textit{Number of Steps} in each task as another independent factor. Mixed Model ANOVA again showed no significant effect for \textit{Technique}, nor its interaction with the \textit{Number of Steps}, although the \textit{Number of Steps} itself were significant for both \textit{Completion Time} and the \textit{Number of App Switches}, which was expected.

\subsubsection{Subjective feedback}

\begin{figure}
  \centering
  \includegraphics[width=\linewidth]{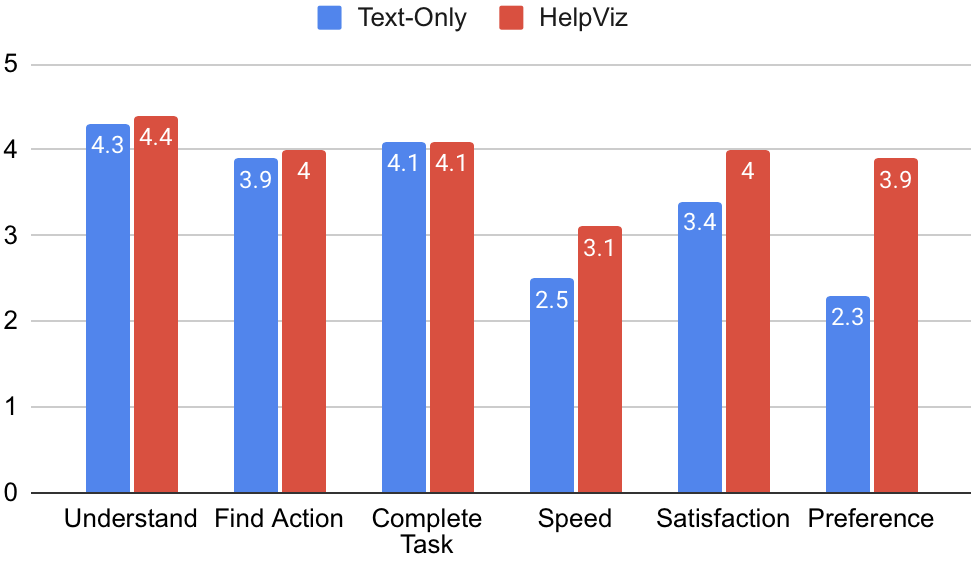}
  \caption{The participants' preferences for each technique.}
  \Description{Bar chart for subjective feedback scores. Overall, the bars for HelpViz appeared taller than for text-only, except for "Complete task", where the bars are of the same height. Participants showed similar preference for ease of understanding, finding actions, and completing tasks. Participants felt that HelpViz was faster, and were more satisfied with HelpViz. Participants also preferred HelpViz overall.}
  \label{fig:subjective}
\end{figure}

We summed subjective scores on each of the Likert items to create an overall scale for each participant, as described in Section~\ref{sec:exp_procedure}. While there was no significance found on completion time and app switch frequency,  participants overall showed preference for HelpViz ($mean=23.5,sd=1.96$) over text instructions ($mean=20.5,sd=2.72$), $t(9)=3.10, p \approx .01$.
For individual items, a Wilcoxon signed-rank test shows that HelpViz scored significantly higher on the \textit{Preference} item ($Z=2.55, p \approx .02, \text{effect size } r=.80$). As shown in Fig.~\ref{fig:subjective}, participants rated HelpViz favorably compared to Text-Only help in general. The overall preference assessment received more positive ratings than other subjective feedback items. This is likely because the Preference item captures the participants' overall assessment about each technique, which is contributed by all aspects of these techniques. Interestingly, although the actual task Completion Time of HelpViz (33 seconds) is on par with that of the Text-Only condition (30 seconds), participants feel HelpViz allow them to finish a task faster, as indicated by the Speed preference in Fig.~\ref{fig:subjective}. We note that the task Completion Time and participants' perception of Speed both showed no statistically significant difference. The faster perceived speed could indicate that the assistance provided by HelpViz made them feel that the tasks took less time.

During the study, several participants expressed their surprise and delight when they found out that the HelpViz tutorial automatically scrolled to their current step (P1, P2, P8). P4 also preferred auto-scroll. None of the participants reported issue with auto-scroll. P2 and P8 praised the quality of the screenshots and auto-generated tutorials. Participants also commented on the benefits of screenshots. 7 out of the 10 participants found the screenshots with UI elements to be helpful. In particular, P2 felt that screenshots make the tutorial easy to follow, and P4, P5, and P8 shared the opinion that screenshots provided expectations for which element to interact with in each step. In addition, 7 out of the 10 participants found the thumbnail to be helpful. P7 favored the screenshots in providing context by showing the neighboring elements and P5 liked seeing an element’s general location within a screen. P5 also mentioned that “images stuck in memory”. P2 and P5 compared the experience of using tutorials with screenshots with text instructions, and stated that text instructions felt harder.

Participants found the ability to fit all instructions into one page for the text instructions was helpful in completing the tasks (9 out of the 10 participants). P7 preferred text for shorter instructions. P6 preferred text instructions in general, claiming that screenshots might not be so valuable for “power users”.

Participants provided insightful feedback and suggestions to improve HelpViz. P1, P3, P8, and P10 all reported that it’s less desirable to use the screenshots when they do not match the app UI, and would like to see screenshots created for their versions of apps. P6 and P9 pointed out that in some cases, such as opening a pop-up menu, it would make more sense to combine menu opening with the next step in the tutorial, because a menu might disappear if they switched out of the app.

\section{Discussion \& Opportunities}

Overall, we received positive feedback from the user study. Our results indicate that while HelpViz does not directly impact the number of operations or total time needed to complete each task, users showed strong preference for HelpViz when compared to text instructions. The screenshots in the tutorials made them easier to follow by allowing the users to have an expectation for the target element and its surrounding elements. Although it is widely known that images are useful in tutorials, previous work shed little light on how visual tutorials that are automatically generated from a batch process are useful, which does not involve any human intervention and manual refinements. Our evaluation contributes new insights that automatically synthesized visual tutorials are preferred by participants, which shows the feasibility of the approach.

In addition, auto-scroll and highlighting of the current step reduced the amount of interaction needed while navigating the tutorial and locating the instruction. Although HelpViz does not seem to significantly affect task execution time in our study, it is preferred by users as the screenshot and auto-scroll features received substantial praise. We hypothesize that as the complexity of the tasks increase, HelpViz can show advantages on time performance as well, which deserves further investigation. Below we describe the limitations of our techniques and identify opportunities that can be readily made to HelpViz.

\textbf{Optimizing Tutorials for Specific Devices \& Granularity.} Participants expressed the desire for more accurate tutorials based on the version of the app on their device. Such tutorials can be easily generated by executing them in emulators running different versions of the Android system and third-party apps. As the entire process is automated, this would require minimal human intervention. Furthermore, certain actions can be grouped together to reduce the length of the tutorial. For example, a menu open action followed by a tap on the menu can be consolidated into one step, which can be achieved by checking the class name and text labels for the menu. Lastly, although none of the participants reported any issue with auto-scroll, screen matching accuracy could be improved by adopting more robust heuristics, such as utilizing other features from the Accessibility API~\cite{sugilite_li17} and adopting the screen equivalence heuristics proposed in~\cite{robust_zhang18}.

\textbf{Supporting Diverse User Groups.} To facilitate power users who prefer text instructions over the images~\cite{chi2012mixt}, HelpViz can present screenshots on demand, or only be displayed for the next step. We suggest that the context-aware capability of HelpViz tutorials will still be valuable in either cases, guiding users to the next step. On the other hand, HelpViz can possibly provide voice guidance from text instructions to support mobile users in need of accessibility support~\cite{10.1007/978-3-030-29381-9_4}. 

\textbf{Generalizing to More Complex Workflows.} Because our tutorial generation pipeline is designed to handle raw, unconstrained text instructions, \eg those directly found on the web, a variety of complex workflows may arise, such as conditional branches and loops. However, these complexities are yet investigated because our experiments focused on PixelHelp, a set of instructions curated from a single website, which contained mostly linear instructions. HelpViz was also constrained by the underlying parsing model, which operated linearly through the instruction text. HelpViz overcomes some of the constraints by supporting branches partially. As typically the branches in a tutorial would transition to different actions for different conditions encountered, the look-ahead mechanism in HelpViz would enable the emulator to skip actions in branches that are not present or invalid. HelpViz does not support loops and would only execute each action in a loop once. These limitations can be addressed by detecting conditionals and loops in instructions in the future.

There are two directions of future work for HelpViz to successfully handle conditional branches: (1) parsing and traversing all possible branches during emulator execution, and (2) matching current encountered state during runtime with recorded states during execution. To solve (1), a more advanced parsing model would be necessary, as well as an infrastructure to set the appropriate emulator environment according to the different branching options in the instruction. To solve (2), HelpViz needs to keep track of the current state during emulator execution, and record the executed steps along with their corresponding states. During runtime, the recorded states can be matched to the states that a user encounters, and tutorials can be shown accordingly. These conditions propose new challenges for the underlying parsing model, and requires additional handling in our pipeline to accommodate these complexities.

Since online tutorials can be complex and diverse, there might be further unexpected situations where the pipeline fails to yield a valid tutorial. Should this happen, as a fallback, our pipeline combines the visual assets from steps that have been successfully executed with the text of the remaining steps.

HelpViz tutorials could be further expanded to desktop workflows, which calls for parsing models tailored to desktop instructions, emulation infrastructure for desktop software, and expanding user action tracking capabilities to a more dynamic environment, where user focus can easily switch between multiple applications.

\section{Conclusion}

HelpViz automates tutorial generation for Android apps by automatically executing a batch of text-based instructions on a cluster of Android emulators, and extracting steps, screenshots, videos snippets, and other metadata that are needed for creating a visual tutorial. By incorporating a previously trained deep model~\cite{li2020mapping}, HelpViz grounds natural language instructions to actions in the emulator---a dynamic environment, and improves the robustness of fully automatic execution by enabling beam search execution and look-ahead. 
A user study with 10 participants indicates that participants reacted positively to HelpViz, particularly when compared to traditional text-only tutorials. HelpViz demonstrates the feasibility to convert a vast number of text instructions on the web to visual tutorials that are adapted to different versions of mobile apps and platforms and hardware configurations.

\section*{Acknowledgements}
We would like to thank anonymous reviewers for their insightful feedback for improving the paper. We also want to thank our user study participants.

\bibliographystyle{ACM-Reference-Format}
\bibliography{ref}

\end{document}